\documentclass[9pt,conference]{IEEEtran}
\usepackage{dcase2026}

\usepackage{bm} 
\usepackage{dcase2026,amsmath,graphicx,url,times,booktabs}

\graphicspath{{figs/}}

\newcommand{\RR}{\mathbb{R}}
\newcommand{\pool}{\mathcal{P}}
\newcommand{\lab}{\mathcal{L}}
\newcommand{\cand}{\mathcal{U}}
\newcommand{\sel}{\mathcal{S}}
\newcommand{\ind}[1]{\mathbb{1}\!\left[#1\right]}

\title{Cover First, Disagree Softly: Rethinking Mismatch-First Active Learning for Frame-Level Audio Classification}

\name{Shiqi Zhang, Tuomas Virtanen}
\address{Audio Research Group, Tampere University, Tampere, Finland\\
\texttt{\{shiqi.zhang, tuomas.virtanen\}@tuni.fi}}

\begin{document}

\maketitle

\begin{abstract}
  Sound event detection relies on frame-level strong labels whose annotation is expensive. Active learning addresses this problem by selecting the audio segments whose labels help the classifier most. One of the prevailing acquisition strategies for this task, mismatch-first farthest-traversal (MFFT), combines the disagreement between two classifiers and the diversity of the selected segments through hard sequential decisions. It selects whole groups of high-disagreement segments first and spreads only the remaining budget by farthest traversal. On two multi-label datasets we show that this design is blind to the similarity among the selected segments and fails under low budgets, with every mismatch-first variant ending below the plain geometric strategy it builds on. We propose mismatch-weighted facility location (MW-FL), which spends the entire budget through a disagreement-weighted coverage objective that penalizes similarity among the selected segments. The disagreement signal from MFFT is used to obtain the nonnegative weights of this facility-location objective, using fixed smoothing without dataset-specific tuning. Experiments across two geometric mechanisms with three ways of using disagreement show that coverage of the selected segments is the dominant factor, hard disagreement gating of selection is harmful on both mechanisms, and soft disagreement weighting helps on top of coverage. MW-FL attains the best area under the learning curve on both datasets.
\end{abstract}

\begin{IEEEkeywords}
  Active learning, sound event detection, frame-level audio classification, submodular maximization, facility location
\end{IEEEkeywords}

\section{Introduction}
\label{sec:intro}

Frame-level audio classification assigns to every short time frame of a recording a multi-label vector of active sound events. It is the core prediction task of sound event detection (SED) \cite{mesaros2021sound} and serves domestic monitoring, environmental noise assessment, and bioacoustic surveys \cite{turpault2019sound, bello2019sonyc, fredianelli2025environmental, stowell2022computational, zhang2025hybrid}. Its bottleneck is the cost of strong labeling, since marking onsets and offsets of every event takes far more annotator time than clip-level tagging \cite{hershey2021benefit}. Pool-based active learning (AL) addresses this by iteratively selecting, under a fixed budget, the audio segments whose labels are expected to help the classifier most \cite{settles2009active, ren2022survey, zhao2017active, kholghi2018active, wang2019active}.

Generic AL offers two families of selection signals, and this paper re-examines how they are combined. Uncertainty- or disagreement-based methods rank candidates by how unsure the current model is \cite{lewis1994sequential, seung1992query, gal2017deep}; in batch acquisition, however, the top of such a ranking is self-similar \cite{kirsch2019batchbald}. Near-identical segments receive the same high score, and labeling them all is inefficient. Diversity methods such as core-set selection spread the batch geometrically \cite{sener2018active, gonzalez1985clustering, hacohen2022active, yehuda2022active} but ignore where the model's predictions might still be wrong. Hybrid strategies therefore combine the two selection signals, and the benefit of such direction has been verified in other classification tasks \cite{brinker2003incorporating, nguyen2004active, ash2020deep}. In frame-level audio classification, however, the prevailing combination strategy remains mismatch-first farthest-traversal (MFFT) \cite{zhao2020active}, most recently applied to bioacoustics \cite{zhang2025hybrid}. It ranks candidates by the disagreement (mismatch) between the classifier and another nearest-neighbor (NN) classifier that predicts the label of the closest labeled segment in embedding space, admits candidates group-by-group in descending mismatch order, and hands only the remaining budget to farthest traversal.

This paper re-examines MFFT under low annotation budgets and reports a negative result. On two frame-level multi-label datasets, DESED \cite{turpault2019sound} and DataSED \cite{fredianelli2025environmental}, MFFT ends below the plain farthest traversal it builds on, and on DESED it even falls below random sampling; the experimental evidence is presented in \cref{sec:exp}.

To locate the failure we factorize MFFT along its two design axes: \emph{how disagreement is used} (not at all, hard mismatch-first gating, or soft weighting) and \emph{which geometric mechanism spends the budget}; we call the latter the \emph{geometric backbone}, as it operates purely on the geometry of the embedding space. In MFFT this backbone is farthest traversal, which repeatedly takes the candidate farthest from everything selected. The factorization exposes two symptoms with a shared root cause. Hard gating admits every member of a high-mismatch group, so the budget concentrates on highly similar segments in a few dense regions, which carry nearly the same information. Farthest traversal, a max--min rule, is an extreme-value statistic; at low budget it drifts toward low-density outliers of the embedding space. Both spend budget through hard decisions, one purely ordinal, one purely geometric; neither accounts for how much \emph{new} information a candidate adds beyond what is already selected, so both are redundancy-blind.

The remedy we argue for is to spend the entire budget through one informativeness-weighted submodular coverage objective. Submodularity means that the marginal coverage gain of a candidate shrinks when similar segments are selected, so the similarity among the selected segments is penalized automatically. Once a dense high-mismatch region is covered, its remaining segments lose most of their marginal value, which removes the gating pathology; an isolated outlier covers little beyond itself, so its contribution is bounded, which removes the traversal pathology. The objective uses fixed smoothing and a data-derived bandwidth without dataset-specific tuning, and has no density term suppressing sparse regions, so genuinely informative sparse regions still receive budget. We instantiate this principle as \emph{mismatch-weighted facility location} (MW-FL).\footnote{The code is publicly available at \url{https://github.com/TioSisai/mismatch-weighted-facility-location}.} It replaces MFFT's traversal backbone with facility location, a coverage objective that favors the subset representing the whole pool, so that every pool segment has a similar segment in the batch and the labeled set, and smooths the nearest-neighbor mismatch signal that MFFT gates on into the nonnegative weights of this objective. Facility location (FL) is well established for selecting representative subsets, with a constant-factor optimality guarantee for greedy maximization \cite{cornuejols1977location, nemhauser1978analysis, wei2015submodularity, kothawade2021similar}; MW-FL only changes its weights and keeps the guarantee.

Our contributions are: (i) a failure diagnosis of mismatch-first acquisition for frame-level audio, (ii) soft mismatch weighting of a facility-location objective that addresses these failures with fixed smoothing and no dataset-specific tuning, and (iii) controlled experimental evidence that separates the two design axes, the geometric backbone and the use of disagreement. For (iii), we compare all six combinations of two geometric backbones, MFFT's farthest traversal and the facility location we introduce, with the three ways of using disagreement, plus random sampling, on two datasets with ten seeds each. The experiments yield three findings. First, the coverage-based geometric backbone, facility location, is the design choice with the largest effect; it clearly beats farthest traversal. Second, hard disagreement gating is harmful on both backbones. Third, soft disagreement weighting helps on top of the coverage backbone. MW-FL ranks first in area under the learning curve (AULC) on both datasets and outperforms all six alternatives, whereas the same weighting on traversal (MW-FT) improves over hard gating yet still fails to beat plain farthest traversal. In these low-budget experiments, soft mismatch weighting improves coverage-based selection, while hard gating reduces performance.

\begin{figure*}[t]
  \centering
  \centerline{\includegraphics[width=\textwidth]{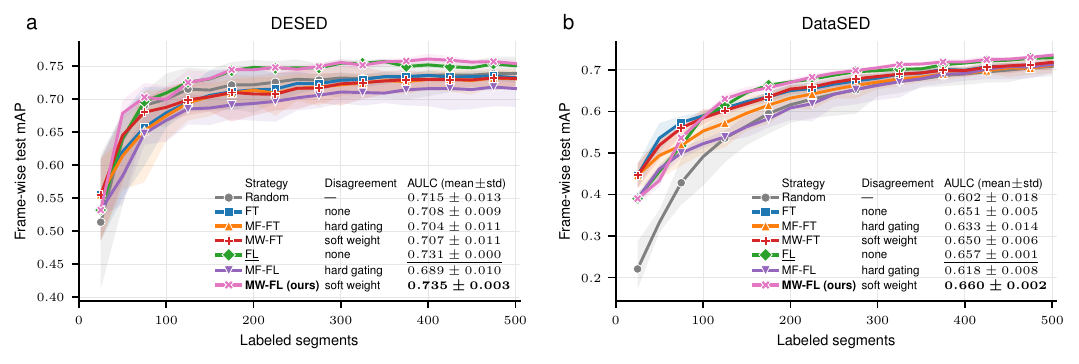}}
  \caption{Test frame-wise mAP versus number of labeled segments on (a) DESED and (b) DataSED. Lines are means over ten seeds, shaded bands min--max ranges; legends report the disagreement handling and the test AULC (mean\(\pm\)std) of every strategy, with \textbf{bold} = best and \underline{underline} = second best per dataset. MW-FL attains the best AULC on both datasets (paired Wilcoxon, \(p{\le}0.006\)), while the mismatch-first variants (MF-FT, MF-FL) trail their geometric backbones (FT, FL) everywhere and fall below random sampling on DESED.}
  \label{fig:curves}
\end{figure*}

\section{Rethinking mismatch-first acquisition}
\label{sec:rethink}

\subsection{Frame-level acquisition setup}
\label{ssec:setup}

Following the setting of MFFT, we consider pool-based AL over a pool \(\pool=\{1,\dots,N\}\) of audio segments with \(C\) event classes. A frozen encoder represents segment \(i\) using a sequence of \(F\) frame embeddings, whose mean is the segment representation \(z_i\in\RR^{D}\). Annotations of a segment represent the frame-wise activities of each of the classes in the segment; the segment-level multi-hot label \(y_i\in\{0,1\}^{C}\) is the frame-wise maximum of these activities. At round \(t\), the labeled set \(\lab_t\subset\pool\) is used to train a frame-level classifier head from scratch. Then, an AL strategy selects a batch \(\sel\) of \(B\) segments from the set of unlabeled segments \(\cand_t=\pool\setminus\lab_t\). Segment-level predictions \(q_i\in[0,1]^{C}\) are the frame-wise maximum of the classifier's frame-wise predictions.

MFFT's disagreement signal compares two predictions for each candidate\cite{zhao2020active,zhang2025hybrid}: the classifier's binarized segment-level prediction and the prediction of a nearest-neighbor (NN) classifier, which outputs the label of the labeled segment closest to \(z_i\) in embedding space, \(\mathrm{nn}(i)\in\lab_t\). It defines the mismatch score
\begin{equation}
  \label{eq:mismatch}
  m_i \;=\; \bigl\lVert \ind{q_i \ge \tfrac{1}{2}} - y_{\mathrm{nn}(i)} \bigr\rVert_1 \;\in\; \{0,\dots,C\},
\end{equation}
where the indicator \(\ind{\cdot}\) binarizes the classifier probabilities at threshold \(\tfrac{1}{2}\) class by class and \(\lVert\cdot\rVert_1\) is the \(\ell_1\) norm; \(m_i\) is thus the Hamming distance between the two multi-hot predictions, i.e., the number of classes on which they disagree. Where the two predictions differ at least one of them is wrong, so querying a segment with \(m_i>0\) can correct an error. The first round has no trained classifier. Traversal-based strategies select an initial set of segments using farthest traversal from a random seed point, and coverage-based strategies run uniform-weight facility location.

\subsection{MFFT: hard gating over a geometric backbone}
\label{ssec:mfft}

MFFT splits the batch in two stages. The first stage proceeds down the occurring mismatch values \(v\), starting from the largest. Each complete group \(G_v=\{i\in\cand_t : m_i=v\}\) of candidates sharing the same mismatch value \(v\) is added to the batch as long as its size \(|G_v|\) fits into the remaining budget; we refer to this all-or-nothing admission as the (mismatch-first) \emph{gate}. The second stage resolves the first group that no longer fits into the remaining budget, the \emph{boundary group}, by farthest traversal \cite{gonzalez1985clustering}. Initialized with the labeled segments and all already selected candidates, it repeatedly picks the candidate whose minimum distance to this base set is largest until the budget is exhausted.

The factorization is natural because the gating stage never depends on which geometric rule resolves the boundary group, and the rule alone is a complete strategy once gating is removed. Crossing the two axes yields six strategies, FT, FL, MF-FT, MF-FL, MW-FT, and MW-FL, where the backbone is farthest traversal (FT) or facility location (FL) and the prefix marks disagreement used as a hard gate (MF-) or as a soft weight (MW-, \cref{sec:method}). MF-FT is MFFT itself with the multi-label mismatch of \cref{eq:mismatch} \cite{zhao2020active, zhang2025hybrid}; the naming merely makes its two components explicit. All six strategies share the same mismatch signal \cref{eq:mismatch} and the same first-round initialization per backbone family, so the experiments in \cref{sec:exp} can attribute any performance difference to the two axes.

\subsection{One root cause, two symptoms}
\label{ssec:diagnosis}

\textbf{Symptom 1: whole-group admission oversamples dense regions.} Groups of segments that enter the batch through the gate bypass the geometric rule entirely; only the boundary group is diversified. High-mismatch segments, however, do not occur in isolation. The classifier and the NN classifier tend to disagree in the same way on a whole cluster of acoustically similar segments at once, so the gate spends several labels on segments that carry nearly the same information. In \cref{fig:selection} (bottom left), MF-FT's 25 selected segments land as a few tight clusters inside two high-mismatch regions, several nearly on top of one another. The more budget the gate consumes, the less the backbone can repair.

\textbf{Symptom 2: farthest traversal is dragged by extremes.} The max--min rule selects whatever lies farthest from everything labeled, and when the budget is a really small fraction of the pool, such candidates are disproportionately low-density outliers far from the bulk of the data (\cref{fig:selection}, FT panel), where labels do little for a frame-wise mAP dominated by the dense majority of the pool. On DESED this alone pulls plain FT below random sampling (\cref{fig:curves}).

\textbf{One root cause.} Both symptoms follow the same decision pattern, an all-or-nothing rule driven by a single criterion (mismatch rank in the gate, minimum distance in the traversal). Under either rule, the score assigned to a candidate never decreases when similar segments enter the batch, so redundancy is invisible to the objective. The fix therefore has to change the objective itself. Under diminishing returns, the defining property of submodular functions \cite{krause2014submodular}, the gain of a candidate shrinks as similar segments are selected, precisely penalizing redundancy.

\section{Cover first, disagree softly}
\label{sec:method}

\subsection{Mismatch-weighted facility location}
\label{ssec:mwfl}

MW-FL selects each batch by greedily maximizing a single weighted coverage objective conditioned on the labeled set. The mismatch \cref{eq:mismatch} is smoothed into a nonnegative weight \(w_i = (m_i + 1)/(C + 1) \in (0,1]\), and the batch maximizes the facility-location value
\begin{equation}
  \label{eq:objective}
  F(\sel) \;=\; \sum_{i\in\cand_t} w_i \max_{j\in\sel\cup\lab_t} k(z_i, z_j),
\end{equation}
where \(k(z,z')=\exp\!\bigl(-\lVert z-z'\rVert^2/(2\sigma^2)\bigr)\) is a radial basis function (RBF) kernel whose bandwidth \(\sigma^2\) is the median of pairwise squared distances among candidates (median heuristic \cite{gretton2012kernel, garreau2017large}). Intuitively, the kernel measures the similarity of two segments; the coverage of candidate \(i\) is its kernel similarity to the closest segment in the batch or in the labeled set, and \(F\) sums this coverage over the pool, weighted by how strongly the two predictors disagree on each candidate. Maximizing \(F\) thus prefers batches that cover every high-disagreement candidate well. Greedy selection runs \(B\) steps, each adding the candidate \(c\) with the largest marginal gain of \cref{eq:objective},
\begin{equation}
  \label{eq:gain}
  \arg\max_{c\in\cand_t\setminus\sel}\; \sum_{i\in\cand_t} w_i \,\bigl[\,k(z_i,z_c) - \mathrm{cov}_i\,\bigr]_+ ,
\end{equation}
where the current coverage \(\mathrm{cov}_i=\max_{j\in\sel\cup\lab_t}k(z_i,z_j)\) of candidate \(i\) is updated per step for \(M=|\cand_t|\) candidates.

The weights, the kernel, and the bandwidth are all determined by the task and the data, without dataset-specific tuning. The weights reuse MFFT's mismatch signal, the \(+1\) smoothing keeps zero-mismatch candidates selectable and reduces the weights to uniform when disagreement vanishes, and the bandwidth comes from the median heuristic. Placing \(\lab_t\) inside the \(\max\) treats the labeled segments as already covering their neighborhoods, so a batch earns value only for covering what the labeled set does not yet cover. With nonnegative weights, \(F\) is monotone and submodular (a weighted facility-location function), so the greedily selected batch is guaranteed to reach at least a \((1{-}1/e)\approx 0.63\) fraction of the objective value of the best possible batch \cite{nemhauser1978analysis, wei2015submodularity, kothawade2021similar}. Two limits locate MW-FL between the design axes: as \(\sigma^2\!\to\!0\) greedy selection degenerates to picking the \(B\) largest weights (pure mismatch ranking), and with uniform weights the objective reduces to plain FL. With these fixed choices, mismatch acts as a soft weight within the coverage objective.

\subsection{Why diminishing returns removes both symptoms}
\label{ssec:why}

\textbf{Within-cluster redundancy control.} The first selected segment inside a dense high-mismatch cluster covers the whole neighborhood, so the marginal gain \cref{eq:gain} of every remaining segment of that cluster drops sharply, regardless of its weight. The budget that hard gating would have spent on further, nearly identical segments is redirected to uncovered regions. Soft weighting preserves exactly the part of the mismatch signal that hard gating overexploits.

\textbf{Outlier contributions are capped.} Selecting an isolated candidate covers little beyond the candidate itself, so it improves \(F\) by at most \(w_c\cdot(1-\mathrm{cov}_c)\le 1\), a contribution capped by its own weight. In farthest traversal, by contrast, the distance of an outlier is the selection criterion itself and grows without limit. Outliers are therefore selected only when no denser region offers more weighted coverage.

\textbf{No density prior.} The weights multiply coverage; nothing divides by density. A sparse region with genuine disagreement retains its full contribution to the objective, so the criterion remains compatible with the observation that sparse, rarely sampled content is often informative, without stacking labels on nearly identical segments.

\subsection{Completing the grid: soft weighting on traversal}
\label{ssec:grid}

To complete the factorial design of \cref{ssec:mfft}, MFFT can be converted to a soft-weighting variant, MW-FT, which reuses the weights \(w_i\) multiplicatively in the max--min rule, picking \(\arg\max_c w_c\, d_{\min}(c)\) with unweighted distance updates. Symmetrically, the remaining variant MF-FL keeps MFFT's gating control flow and resolves the boundary group by uniform-weight facility location whose bandwidth is estimated on the full candidate pool, so the kernels of all FL variants are identical and \cref{sec:exp} can attribute every gap to one axis at a time.

\begin{figure*}[t]
  \centering
  \centerline{\includegraphics[width=\textwidth]{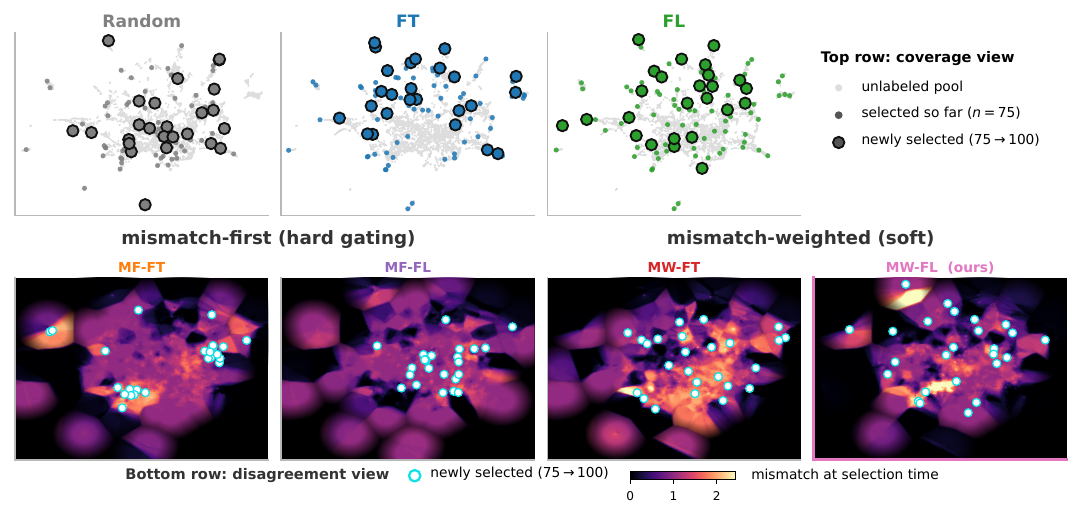}}
  \caption{Segments selected in one acquisition round (DataSED, round \(75{\to}100\), seed 0, shared UMAP projection \cite{mcinnes2018umap}; marker and color coding are given in the legend). \textbf{Top, coverage view:} cumulative selections of the disagreement-free strategies; FT concentrates on sparse regions far from the data bulk, FL spreads evenly across the dense regions. \textbf{Bottom, disagreement view:} the 25 new selections of the disagreement-aware strategies on the mismatch field at selection time (the previous round's \(m\) smoothed over the pool). Hard gating (MF-FT, MF-FL) stacks its selections inside a few high-mismatch regions; soft weighting (MW-FT, MW-FL) spreads the same budget across the same structure.}
  \label{fig:selection}
\end{figure*}

\section{Experiments}
\label{sec:exp}

\subsection{Setup}
\label{ssec:exp_setup}

\textbf{Datasets.} \emph{DESED} \cite{turpault2019sound}: the pool is the 10\,000 synthetic 10-s soundscapes of the DCASE 2021 Task 4 synthetic training set \cite{ronchini2021impact}, validation uses the corresponding 2\,500 synthetic soundscapes, and testing uses the 693 segments of the public evaluation set of real recordings; \(C{=}10\). \emph{DataSED} \cite{fredianelli2025environmental}: real-world environmental noise recordings with polyphonic strong labels, segmented into 6\,808 10-s segments and split 8:1:1 with label stratification into a pool of 5\,446 and 681 segments each for validation and testing; \(C{=}22\).

\textbf{Model and protocol.} A frozen PANNs Cnn14 encoder \cite{kong2020panns} yields 31 frame embeddings (\(D{=}2048\)) per segment. The classifier is a two-layer MLP (\(2048{\to}512{\to}C\)) applied frame-wise and trained with binary cross-entropy from scratch every round (25 epochs, Adam, learning rate \(10^{-3}\), batch size 16, 20\% linear warmup then cosine decay), keeping the epoch with the best validation frame-wise mAP. Acquisition proceeds in 20 rounds of \(B{=}25\) segments up to 500 labels (5.0\% of the DESED pool, 9.2\% of DataSED); the first round is the model-free cold start of \cref{ssec:setup}. All disagreement-aware strategies compute the mismatch \cref{eq:mismatch} from the same NN classifier, and every configuration runs with ten seeds.

\textbf{Metrics.} We report frame-wise macro mAP on the test split, and summarize each learning curve by its normalized AULC, the trapezoidal area under the mAP-versus-labels curve divided by the budget span (same scale as mAP). Strategies are compared with paired Wilcoxon signed-rank tests \cite{wilcoxon1945individual} across seeds.

\subsection{Main comparison}
\label{ssec:main}

\Cref{fig:curves} gives the main result. MW-FL attains the best test AULC on both datasets (\(0.735\) on DESED, \(0.660\) on DataSED) and beats each of the six alternatives under the paired Wilcoxon test (\(p\le 0.006\), \({\ge}9\) of 10 seeds won in every comparison). Its difference to the strongest baseline, plain FL, is small but consistent (\(+0.004\) mean AULC on both datasets). The differences to the mismatch-first family are large (\(+0.046\) and \(+0.042\) relative to MF-FL, \(+0.031\) and \(+0.027\) relative to MF-FT). On DESED, MW-FL reaches a higher test mAP with 200 labels (\(0.744\)) than random sampling with the full 500-label budget (\(0.739\)).

The negative result is equally clear. On both datasets each mismatch-first variant sits below the plain backbone it gates, MF-FT below FT (\(-0.004\) DESED, \(-0.018\) DataSED) and MF-FL below FL (\(-0.042\) and \(-0.038\)). On DESED, where random sampling is strong, both MF variants also fall below Random; MF-FL trails it by \(0.026\) AULC despite building on FL. Hard gating cancels out most of its backbone's gains.

\subsection{Isolating the two axes}
\label{ssec:axes}

Reading the AULC legends of \cref{fig:curves} along the factorial axes yields three findings.

\textbf{(i) The coverage backbone is the dominant factor.} With no or soft disagreement, FL beats FT on both datasets (none: \(+0.023\) DESED, \(+0.005\) DataSED; soft: \(+0.028\), \(+0.010\)), and FL alone is already the second-best strategy everywhere. At low budget farthest traversal spends a visible share of its batch on outliers (\cref{fig:selection}); on DESED plain FT stays below Random over the entire budget range. Coverage also brings stability. The standard deviations of the FL family are several times smaller because greedy coverage selection is deterministic given the embeddings and weights.

\textbf{(ii) Hard gating is harmful on both backbones.} The hard-gating variants MF-FT and MF-FL are the worst within their backbone families on both datasets. Gating even reverses the ranking of the two backbones; MF-FL falls below MF-FT everywhere. The better the geometric rule covers the dense core of the boundary group, the more its selections overlap with the already admitted groups, so a stronger backbone amplifies rather than repairs the gate's redundancy.

\textbf{(iii) Soft weighting helps on top of coverage.} MW-FL improves on FL on both datasets, so the mismatch signal does carry information that pure coverage misses. The same weighting on traversal recovers most of the gating damage (MW-FT vs.\ MF-FT: \(+0.004\) DESED, \(+0.017\) DataSED) yet still fails to beat plain FT. Disagreement helps only as a soft modulation of an objective that already penalizes redundancy; in a max--min rule it merely changes which extreme points are picked.

\subsection{Where one round's budget goes}
\label{ssec:budget}

The two failure modes appear in \cref{fig:selection} where the diagnosis places them. MF-FT stacks its 25 selections into two bright high-mismatch regions (the whole-group admissions of \cref{ssec:diagnosis}) and MF-FL concentrates them in the dense center, while FT drifts to sparse regions and FL covers the populated part of the embedding space evenly. The soft strategies land on the same bright structure but spread across it. MW-FL's selections sit on bright regions while keeping distance from one another.

\section{Conclusion}
\label{sec:conclusion}

We revisited mismatch-first farthest-traversal, the prevailing acquisition strategy for frame-level audio, and traced its low-budget failure to one root cause, hard and redundancy-blind budget spending. We proposed mismatch-weighted facility location (MW-FL) to address this failure. It spends the entire budget by greedily maximizing one submodular coverage objective over the pool, reusing MFFT's mismatch signal with fixed smoothing and no dataset-specific tuning. Its diminishing returns remove both symptoms, as covered clusters stop attracting budget and an outlier contributes at most its own weight. MW-FL ranks first in AULC on both datasets. In the low-budget settings, the controlled evidence shows that coverage is the dominant factor, hard gating harms both tested backbones, and soft mismatch weighting improves coverage. These findings support covering the pool first and using mismatch as a soft weight. Higher budgets, other informativeness signals, and broader baseline comparisons remain for future work.

\clearpage
\IEEEtriggeratref{19}
\bibliographystyle{IEEEtran}
\bibliography{refs}

@string{icassp = "Proc. ICASSP"}

@string{dcase = "Proc. DCASE Workshop"}

@string{ieee-acm-taslp = "IEEE/ACM Trans. Audio, Speech, Lang. Process."}

@string{icml = "Proc. ICML"}

@string{iclr = "Proc. ICLR"}

@string{neurips = "Proc. NeurIPS"}

@string{jmlr = "JMLR"}

@string{eusipco = "Proc. EUSIPCO"}

@string{sigir = "Proc. ACM SIGIR"}

@article{zhao2020active,
  author  = {Zhao, Shuyang and Heittola, Toni and Virtanen, Tuomas},
  title   = {Active learning for sound event detection},
  journal = ieee-acm-taslp,
  volume  = {28},
  pages   = {2895--2905},
  year    = {2020}
}

@inproceedings{zhao2017active,
  author    = {Zhao, Shuyang and Heittola, Toni and Virtanen, Tuomas},
  title     = {Active learning for sound event classification by clustering unlabeled data},
  booktitle = icassp,
  pages     = {751--755},
  year      = {2017}
}

@inproceedings{zhang2025hybrid,
  author    = {Zhang, Shiqi and Virtanen, Tuomas},
  title     = {Hybrid disagreement-diversity active learning for bioacoustic sound event detection},
  booktitle = eusipco,
  year      = {2025}
}

@article{kholghi2018active,
  author  = {Kholghi, Mahnoosh and Phillips, Yvonne and Towsey, Michael and Sitbon, Laurianne and Roe, Paul},
  title   = {Active learning for classifying long-duration audio recordings of the environment},
  journal = {Methods in Ecology and Evolution},
  volume  = {9},
  number  = {9},
  pages   = {1948--1958},
  year    = {2018}
}

@inproceedings{wang2019active,
  author    = {Wang, Yu and Mendez Mendez, Ana Elisa and Cartwright, Mark and Bello, Juan Pablo},
  title     = {Active learning for efficient audio annotation and classification with a large amount of unlabeled data},
  booktitle = icassp,
  pages     = {880--884},
  year      = {2019}
}

@techreport{settles2009active,
  author      = {Settles, Burr},
  title       = {Active learning literature survey},
  institution = {University of Wisconsin--Madison},
  number      = {Computer Sciences Technical Report 1648},
  year        = {2009}
}

@inproceedings{lewis1994sequential,
  author    = {Lewis, David D. and Gale, William A.},
  title     = {A sequential algorithm for training text classifiers},
  booktitle = sigir,
  pages     = {3--12},
  year      = {1994}
}

@inproceedings{sener2018active,
  author    = {Sener, Ozan and Savarese, Silvio},
  title     = {Active learning for convolutional neural networks: A core-set approach},
  booktitle = iclr,
  year      = {2018}
}

@inproceedings{ash2020deep,
  author    = {Ash, Jordan T. and Zhang, Chicheng and Krishnamurthy, Akshay and Langford, John and Agarwal, Alekh},
  title     = {Deep batch active learning by diverse, uncertain gradient lower bounds},
  booktitle = iclr,
  year      = {2020}
}

@inproceedings{seung1992query,
  author    = {Seung, H. Sebastian and Opper, Manfred and Sompolinsky, Haim},
  title     = {Query by committee},
  booktitle = {Proc. COLT},
  pages     = {287--294},
  year      = {1992}
}

@inproceedings{gal2017deep,
  author    = {Gal, Yarin and Islam, Riashat and Ghahramani, Zoubin},
  title     = {Deep {Bayesian} active learning with image data},
  booktitle = icml,
  pages     = {1183--1192},
  year      = {2017}
}

@inproceedings{kirsch2019batchbald,
  author    = {Kirsch, Andreas and van Amersfoort, Joost and Gal, Yarin},
  title     = {{BatchBALD}: Efficient and diverse batch acquisition for deep {Bayesian} active learning},
  booktitle = neurips,
  pages     = {7024--7035},
  year      = {2019}
}

@inproceedings{brinker2003incorporating,
  author    = {Brinker, Klaus},
  title     = {Incorporating diversity in active learning with support vector machines},
  booktitle = icml,
  pages     = {59--66},
  year      = {2003}
}

@inproceedings{nguyen2004active,
  author    = {Nguyen, Hieu Tat and Smeulders, Arnold W. M.},
  title     = {Active learning using pre-clustering},
  booktitle = icml,
  year      = {2004}
}

@article{ren2022survey,
  author  = {Ren, Pengzhen and Xiao, Yun and Chang, Xiaojun and Huang, Po-Yao and Li, Zhihui and Gupta, Brij B. and Chen, Xiaojiang and Wang, Xin},
  title   = {A survey of deep active learning},
  journal = {ACM Comput. Surv.},
  volume  = {54},
  number  = {9},
  pages   = {180:1--180:40},
  year    = {2022}
}

@inproceedings{hacohen2022active,
  author    = {Hacohen, Guy and Dekel, Avihu and Weinshall, Daphna},
  title     = {Active learning on a budget: Opposite strategies suit high and low budgets},
  booktitle = icml,
  pages     = {8175--8195},
  year      = {2022}
}

@inproceedings{yehuda2022active,
  author    = {Yehuda, Ofer and Dekel, Avihu and Hacohen, Guy and Weinshall, Daphna},
  title     = {Active learning through a covering lens},
  booktitle = neurips,
  pages     = {22354--22367},
  year      = {2022}
}

@inproceedings{wei2015submodularity,
  author    = {Wei, Kai and Iyer, Rishabh and Bilmes, Jeff},
  title     = {Submodularity in data subset selection and active learning},
  booktitle = icml,
  pages     = {1954--1963},
  year      = {2015}
}

@inproceedings{kothawade2021similar,
  author    = {Kothawade, Suraj and Beck, Nathan and Killamsetty, Krishnateja and Iyer, Rishabh},
  title     = {{SIMILAR}: Submodular information measures based active learning in realistic scenarios},
  booktitle = neurips,
  year      = {2021}
}

@article{nemhauser1978analysis,
  author  = {Nemhauser, George L. and Wolsey, Laurence A. and Fisher, Marshall L.},
  title   = {An analysis of approximations for maximizing submodular set functions---{I}},
  journal = {Mathematical Programming},
  volume  = {14},
  pages   = {265--294},
  year    = {1978}
}

@incollection{krause2014submodular,
  author    = {Krause, Andreas and Golovin, Daniel},
  title     = {Submodular function maximization},
  booktitle = {Tractability: Practical Approaches to Hard Problems},
  publisher = {Cambridge University Press},
  pages     = {71--104},
  year      = {2014}
}

@article{gonzalez1985clustering,
  author  = {Gonzalez, Teofilo F.},
  title   = {Clustering to minimize the maximum intercluster distance},
  journal = {Theoretical Computer Science},
  volume  = {38},
  pages   = {293--306},
  year    = {1985}
}

@article{cornuejols1977location,
  author  = {Cornu{\'e}jols, G{\'e}rard and Fisher, Marshall L. and Nemhauser, George L.},
  title   = {Location of bank accounts to optimize float: An analytic study of exact and approximate algorithms},
  journal = {Management Science},
  volume  = {23},
  number  = {8},
  pages   = {789--810},
  year    = {1977}
}

@article{gretton2012kernel,
  author  = {Gretton, Arthur and Borgwardt, Karsten M. and Rasch, Malte J. and Sch{\"o}lkopf, Bernhard and Smola, Alexander},
  title   = {A kernel two-sample test},
  journal = jmlr,
  volume  = {13},
  pages   = {723--773},
  year    = {2012}
}

@article{garreau2017large,
  author  = {Garreau, Damien and Jitkrittum, Wittawat and Kanagawa, Motonobu},
  title   = {Large sample analysis of the median heuristic},
  journal = {arXiv preprint arXiv:1707.07269},
  year    = {2017}
}

@inproceedings{turpault2019sound,
  author    = {Turpault, Nicolas and Serizel, Romain and Salamon, Justin and Shah, Ankit Parag},
  title     = {Sound event detection in domestic environments with weakly labeled data and soundscape synthesis},
  booktitle = dcase,
  pages     = {253--257},
  year      = {2019}
}

@inproceedings{ronchini2021impact,
  author    = {Ronchini, Francesca and Serizel, Romain and Turpault, Nicolas and Cornell, Samuele},
  title     = {The impact of non-target events in synthetic soundscapes for sound event detection},
  booktitle = dcase,
  pages     = {115--119},
  year      = {2021}
}

@article{fredianelli2025environmental,
  author  = {Fredianelli, Luca and Artuso, Francesco and Pompei, Geremia and Licitra, Gaetano and Iannace, Gino and Akbaba, Andac},
  title   = {Environmental noise dataset for sound event classification and detection},
  journal = {Scientific Data},
  volume  = {12},
  number  = {1712},
  year    = {2025}
}

@article{kong2020panns,
  author  = {Kong, Qiuqiang and Cao, Yin and Iqbal, Turab and Wang, Yuxuan and Wang, Wenwu and Plumbley, Mark D.},
  title   = {{PANNs}: Large-scale pretrained audio neural networks for audio pattern recognition},
  journal = ieee-acm-taslp,
  volume  = {28},
  pages   = {2880--2894},
  year    = {2020}
}

@article{mcinnes2018umap,
  author  = {McInnes, Leland and Healy, John and Melville, James},
  title   = {{UMAP}: Uniform manifold approximation and projection for dimension reduction},
  journal = {arXiv preprint arXiv:1802.03426},
  year    = {2018}
}

@article{mesaros2021sound,
  author  = {Mesaros, Annamaria and Heittola, Toni and Virtanen, Tuomas and Plumbley, Mark D.},
  title   = {Sound event detection: A tutorial},
  journal = {IEEE Signal Process. Mag.},
  volume  = {38},
  number  = {5},
  pages   = {67--83},
  year    = {2021}
}

@article{bello2019sonyc,
  author  = {Bello, Juan Pablo and Silva, Cl{\'a}udio and Nov, Oded and DuBois, R. Luke and Arora, Anish and Salamon, Justin and Mydlarz, Charles and Doraiswamy, Harish},
  title   = {{SONYC}: A system for monitoring, analyzing, and mitigating urban noise pollution},
  journal = {Commun. ACM},
  volume  = {62},
  number  = {2},
  pages   = {68--77},
  year    = {2019}
}

@article{stowell2022computational,
  author  = {Stowell, Dan},
  title   = {Computational bioacoustics with deep learning: a review and roadmap},
  journal = {PeerJ},
  volume  = {10},
  pages   = {e13152},
  year    = {2022}
}

@inproceedings{hershey2021benefit,
  author    = {Hershey, Shawn and Ellis, Daniel P. W. and Fonseca, Eduardo and Jansen, Aren and Liu, Caroline and Moore, R. Channing and Plakal, Manoj},
  title     = {The benefit of temporally-strong labels in audio event classification},
  booktitle = icassp,
  pages     = {366--370},
  year      = {2021}
}

@article{wilcoxon1945individual,
  author  = {Wilcoxon, Frank},
  title   = {Individual comparisons by ranking methods},
  journal = {Biometrics Bulletin},
  volume  = {1},
  number  = {6},
  pages   = {80--83},
  year    = {1945}
}

\end{document}